\documentclass{aa}
\usepackage{graphicx}
\def\spose#1{\hbox to 0pt{#1\hss}}
\def\lta{\mathrel{\spose{\lower 3pt\hbox{$\mathchar"218$}}
     \raise 2.0pt\hbox{$\mathchar"13C$}}}
\def\gta{\mathrel{\spose{\lower 3pt\hbox{$\mathchar"218$}}
     \raise 2.0pt\hbox{$\mathchar"13E$}}}

\newcommand{\etal}{et al. }

\begin{document}
\title{Evolutionary Synthesis Models for the Formation of S0 Galaxies in
  Clusters} 
\subtitle{}
\author{J. Bicker 
  \and U. Fritze -- v. Alvensleben 
  \and K. J. Fricke}
\institute{Universit\"atssternwarte G\"ottingen,
  Geismarlandstr. 11, 37083 G\"otingen, Germany\\
  email:jbicker@uni-sw.gwdg.de} 
\authorrunning{J. Bicker \etal}
\titlerunning{S0 Galaxies in Clusters}
\offprints{J.~Bicker, \email{jbicker@uni-sw.gwdg.de}}
\date{Submitted December 17, 2001, accepted March 18, 2002}

\abstract{Rich galaxy clusters in the local Universe show a large population of S0
  galaxies ($\sim 40$\% of all luminous galaxies). With increasing redshift the fraction
  of this S0 galaxy population is observed to strongly decrease (e.g. by a factor $\sim
  2-3$ to ${\rm z = 0.5}$) in favor of the spiral galaxy fraction while the number of
  bright ellipticals does not seem to change.  The infalling field galaxy population that
  successively builds up the cluster also is spiral rich and S0 poor. It has hence been
  suspected that galaxy transformation processes, either due to galaxy -- galaxy or to
  galaxy -- ICM interactions, are responsible for this change. Complementing dynamical and
  morphological studies, we use evolutionary synthesis models describing various possible
  effects of those interactions on the star formation rates of the infalling spirals. We
  study the effects of starbursts of various strengths as well as of the truncation of
  star formation on the color and luminosity evolution of model galaxies of various
  spectral types. Comparison with observed properties of the local S0 galaxy population is
  used to constrain possible S0 formation mechanisms. We find that star formation
  truncation in spiral galaxies earlier than Sd-type, if occurring not too long ago, as
  well as starbursts more  than 3 Gyr ago and followed by complete star formation
  extinction in spirals -- again earlier than Sd-- may well account for the observed
  average S0 luminosities and colors. Late-type galaxies (Sd), even after a strong burst,
  remain either  too blue or too faint. Our results are in agreement with studies of
  spectral features of cluster S0s but allow for stronger constraints. 
  }

\maketitle

\section{Introduction}

The galaxy populations in rich clusters are strikingly different at
intermediate and high redshift from what they are in the local
Universe.  As early as 1978 Butcher \& Oemler noticed a large fraction
of blue galaxies in the centers of distant clusters in contrast to the
situation in the centers of nearby rich galaxy clusters where almost
exclusively red and passive early-type galaxies are found. While the
early observations referred to color or spectral classification of
galaxies, HST has opened up the possibility for direct morphological
classifications of distant galaxies.

While the composition of the bright field galaxy population in terms of morphological
types (E, S0 and Spirals), does not seem to change significantly up to $z\ge 0.5$, that of
the populations in rich galaxy clusters does. S0 galaxies, in particular, constitute $\sim
40$\% of the local cluster population while 10 rich clusters at $z\sim 0.5$ have been shown
to only contain $\sim 15$\% S0 galaxies (Dressler 1980, Dressler \& Gunn 1992, Dressler
\etal 1997, 1999, Smail \etal 1997). This trend has been confirmed by additional studies of
12 clusters in the redshift range between $z\sim 0.1$ and $0.3$ by Couch \etal 1996 and
Fasano \etal 2000. The proportion of elliptical galaxies in the central regions of rich
clusters does not seem to change significantly with redshift while that of the spirals does
in a sense opposite to that of the S0s.

Structure formation scenarios show that even in the local Universe
galaxy clusters are still in their process of formation by
continuously accreting galaxies from the spiral-rich field
environment.

Various processes have been discussed that might act to transform
infalling star-forming field spirals into the passive S0 galaxy
population seen inside nearby rich clusters like eg. Coma. This
transformation has two aspects: the morphological transformation and
the truncation of the star formation rate (SFR). The timescales for
both processes may or may not be different.

Spiral-spiral mergers are known from dynamical \& spectrophotometric
modeling and from the multi-frequently observations of nearby
examples (NGC 4038/39, NGC7252) to lead to spheroidal remnants of type
S0/E, excite sometimes very strong bursts of SF during the
interaction, consuming much of the gas, eventually expelling the rest,
leaving a gas-poor remnant with low or vanishing SFR and colors \&
spectral features evolving from those of a typical postburst galaxy
with strong Balmer absorption lines to those of ``normal'' E/S0
galaxies (Toomre \& Toomre 1972, Barnes \& Hernquist 1996, Fritze --
v. Alvensleben \& Gerhard 1994, Schweizer 1999).

The high velocity dispersions of galaxies in rich clusters today,
however, are unfavorable to galaxy merging. During the early stages
of cluster formation or within galaxy groups falling into a cluster,
However, merging may play an important role.  Harassment has been
shown by Moore \etal (1996, 1998) to be efficient in stripping stars
from spiral disks during their high speed encounters with other
cluster members and/or inhomogeneities in the cluster potential.
However, only for massive bulge-dominated infalling galaxies S0-type
remnants are expected. Late-type or low surface brightness galaxies
will rather be shred down to dSphs or dEs.

Beyond these galaxy-galaxy and galaxy-potential interactions a third
type of interaction is shown to be important, both observationally and
from theoretically modeling: that of a galaxy with the dense hot ICM
filling the central regions of clusters -- at least in the local
Universe -- as seen in X-ray observations.  As this hot ICM can reach
densities up to a factor 100 higher than those of the ISM in spiral
galaxies it will cause strong shocks in the ISM of spiral galaxies
falling towards the cluster center as well as efficient ram pressure
stripping or sweeping of its ISM (Gunn \& Gott 1972, Abadi \etal 1999,
Fujita 2001).  Quilis \etal 2000 present 3D hydrodynamic simulations
with good shock resolution showing that ram pressure together with
turbulent and viscous stripping can remove a galaxy's HI content and
hence extinguish its SF within $\sim 10^8$yr. Preexisting molecular
clouds and the strong shocks induced within the ISM may give a rise to
starburst during this process.

Both hydrodynamic models and observations, showing that all of the
starburst and most of the post-burst galaxies in clusters are spirals,
indicate that changes in SFR take place on shorter timescales ($\le
10^8$yr) than morphological transformations ($\sim 10^9$yr).

Observationally, the hydrogen deficiency of spiral galaxies near or
around the central regions of dynamically relaxed clusters and their
asymmetric or displaced HI distributions are direct evidence of the
effect of the hot ICM (Cayatte \etal 1990, Bravo-Alfaro \etal 2000,
Poggianti \& van Gorkom 2001). The spectacular $\textrm{H}_\alpha$
emission of some Virgo and A1367 galaxies as they are stripped
confirms the starbursts predicted to be induced during this process
(Gavazzi \etal 1995, Kenney \& Koopman 1999).

A delay of 1-2 Gyr between SF truncation and the morphological
transformation in a infalling spiral is consistent with the lack of
blue S0s in distant clusters as well as with the large number of
passive galaxies with late-type morphologies (Poggianti \etal 1999,
Kodama \& Smail 2001, Smail \etal 2001).  Ziegler 2000 and Smail \etal
2001 find that while luminous E/S0s show optical-NIR colors and
spectral line strengths consistent with old systems and passive
evolution since $z>0.5$, the fainter E/S0s ($\le 0.1 {\rm
  L_K^{\ast}}$) show a large spread in optical and optical-NIR colors
with $\sim30$\% of the S0s in A2218 ($z=0.17$) having
luminosity-weighted ages $\le 5$ Gyr, i.e. they were actively forming
stars at $z\le 0.5$. This proportion of young S0s is consistent with
the rate of evolution in the fraction of S0s in clusters over the
redshift range from $z\sim 0.5$ to $z=0$ (Dressler \etal 1997).
Similar young components are reported for 3 other S0 galaxies at the
periphery of the Fornax cluster by Kuntschner 2000.

While most of these studies have concentrated on the inner regions of
rich clusters, both at high and low redshift, it is clear that, in
fact, the evolution of galaxy clusters themselves, in terms of
richness, degree of relaxation, and content of ICM have to be included
into any analysis of the evolution of their galaxy populations. Very
first attempts of this kind are presented by Balogh \etal 2000.

In this paper we take a much more modest approach. We use evolutionary
synthesis models that were earlier proven to correctly describe the
evolution of undisturbed field galaxies of various spectral types over
significant look-back times and we artificially modify their SFRs at
different evolutionary stages as suggested to happen when these
galaxies are assumed to fall into a cluster. I. e. we induce a burst
of SF and/or truncate the SF on timescales as indicated by
observations and/or hydrodynamical simulations. We then investigate
the subsequent evolution of our model galaxies in terms of optical and
optical-NIR colors and luminosities and compare with the corresponding
data for the local S0 cluster galaxy population to try and constrain
possible photometric evolutionary paths from field spirals to cluster
S0s. It may be of historical interest that already in 1975 Biermann \&
Tinsley presented an early version of an evolutionary synthesis study,
conducted mainly in G\"ottingen, on the effect of SF truncation in
spirals in comparison with cluster S0s, on the basis of stellar input
physics (main sequence evolutionary tracks for solar metallicity
stars) and computing facilities available at that time.

We present the basic principles of our models in Sect.~2, discuss
the results from a grid of models for different spiral types
experiencing different changes of their SFR at various evolutionary
stages in Sect.~3, and compare to observational properties of local
S0 galaxies in Sect.~4. Sect.~5 summarizes our conclusions,
compares to results from other groups studying spectral features, and
gives an outlook to future developments of our models an observations.

\section{The spectrophotometric evolutionary model}

\subsection{Undisturbed galaxies}

The evolutionary synthesis model we use is based on Tinsley's
equations, which describe the time evolution of the gas mass, the
stellar mass, and the ISM metallicity on the basis of a given initial
mass function (IMF) and a star formation rate in its time evolution
specific for each galaxy type (e.g. Tinsley 1980). At the same time
our code follows the detailed evolution of the stellar population
using a complete set of stellar evolutionary tracks including all
relevant stellar evolutionary phases, as given by the Geneva group
(cf. Weilbacher \etal 2000 for details), stellar yields and lifetimes.
It is a 1-zone model for the description of global galaxy properties.
Starting from a homogeneous gas cloud of initial mass and chemical
composition our program calculates the distribution of the stellar
population over the HRD at every timestep.  The model assumes ideal
gas mixing and no in- or outflow of gas, but accounts for the finite
lifetimes of stars of different masses. The various spiral galaxy
types are described by conventional global star formation rates.
For Sa, Sb, and Sc galaxies we assume SFRs which are linear functions of the evolving
gas content G(t), $\psi(t)\sim G(t)$ (Schmidt 1959, Kennicutt 1998). The Sd model has a
constant SFR. These star formation rates are close to those empirically determined by
Sandage (1986), and also used by Bruzual \& Charlot (1993) and Guiderdoni \&
Rocca-Volmerange (1987). The IMF is taken from Scalo (1986):
\begin{eqnarray}
  {\rm \phi(m)dm \propto m^{-(1+x)}dm}
\end{eqnarray}
with lower and upper mass limits ${\rm m_l=0.15~M_\odot}$ and ${\rm
  m_u=85~M_\odot}$. The three different slopes are ${\rm x_1=0.25}$
for ${\rm m_l \le m < 1}$, ${\rm x_2=1.35}$ for ${\rm 1 \le m <2}$ and
${\rm x_3=2.00}$ for ${\rm 2 \le m \le m_u}$. To account for dark
matter in the form of brown dwarfs or planets the normalization
\begin{eqnarray}
{\rm \int_{m_l}^{m_u} \phi(m) m dm =0.5}
\end{eqnarray}(Bahcall et al. 1992)
is used which, after a Hubble time, gives mass-to-light ratios
for our model galaxies in agreement with observations. We use the
tracks for a metallicity of  $\frac{1}{2}\textrm{Z}_\odot$, which give
a good description for stars in spiral galaxies (Rocha-Pinto \&
Maciel 1998).

The model calculates the time evolution of luminosities in the Johnson
UBVRIJHK bands and the corresponding colors.

The SFRs chosen above were carefully checked to yield agreement after 12 Gyr -- the
present galaxy age in our cosmology-- with average observed integrated colors from U
through K for the respective galaxy types (RC3, Buta et al. 1994) At an age of 12 Gyr
(today), the total B-band luminosities of the different undisturbed galaxy models are
scaled to the average ${\rm \langle M_B\rangle}$ from Gauss-fits to observed type-specific
luminosity functions of the respective galaxy types in Virgo  as derived by Sandage \etal
(1985). This scaling also is applied to the total, stellar and gaseous masses of the model
galaxies.

\subsection{Interacting galaxies}

To model interacting galaxies we use two different scenarios. Starbursts are used to
describe merging galaxies or galaxies with high SFRs triggered by shock waves in the ISM of
a galaxy falling into the hot dense ICM. Star formation truncation, on the other hand, is
used to model galaxies which lose their gas by tidal interaction or by ram pressure
stripping when falling into a cluster.

In case of the truncation we simply stop star formation at a given
point in time. A burst is described by a sudden increase of the SFR
followed by an exponential decline either to zero or to some given
value ${\rm \psi_f}$ assumed to be constant thereafter.
\begin{itemize}
\item ${\rm \psi_{max}}$: max. SFR at the beginning of the burst, i.e. at 
${\rm t_{burst}}$
\item ${\rm \tau_{burst}}$: exponential decline timescale of the burst SFR, typically ${\rm \sim 10^8~ yr}$
\item ${\rm \psi_f = 0 ~or~=1.5~M_{\odot}yr^{-1}}$: eventual low level constant SFR after the burst
\end{itemize}
The strength of the burst b is defined by the fraction of gas available
at the time when the burst starts ${\rm G(t_{burst})}$, which is
consumed for star formation during the burst (${\rm \Delta G}$):
\begin{eqnarray}
  {\rm b:=\frac{\Delta G}{G(t_{burst})}}
\end{eqnarray}

Another common definition of the burst strength is the fraction of stars built in the
burst $(\Delta S)$ as compared to the total stellar mass of the galaxy by today $(S(t_0))$:

\begin{eqnarray}
  {\rm b_*:=\frac{\Delta S}{S(t_0)}}
\end{eqnarray}

\section{Model Results}
With the code described above we compute a grid of models for spiral
galaxies falling into a cluster and having their SFRs affected by
starbursts and/or star formation truncation as described above. The
different parameters are:
\begin{itemize}
\item Spectral types Sa, Sb, Sc, and Sd of the infalling galaxies -- defined by their respective SF histories
\item evolutionary age of the galaxies at the onset of the burst or the 
star formation truncation: 3, 6, and 9 Gyr
\end{itemize}
and in case of a starburst 
\begin{itemize}
\item burst strength b in terms of gas consumption: 70\%, 50\%, and
  30\% (strong, intermediate and weak bursts) 
\item SFR after the burst ${\rm \psi_f=1.5~M_{\odot}
yr^{-1}}$ or ${\rm \psi_f=0}$.
\end{itemize}
\begin{figure}[h]
\includegraphics[width=\columnwidth,angle=0]{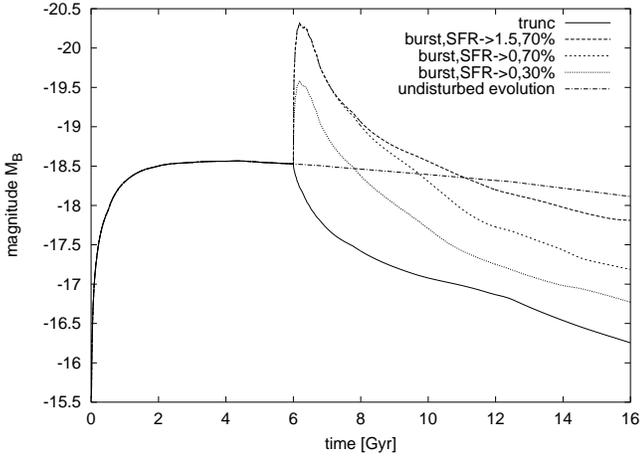}
\caption{Time evolution of B-band luminosities of Sc galaxies in
  different interaction scenarios with the SFR affected by strong and weak bursts and SF
  truncation after 6 Gyr of undisturbed evolution.}
\label{m01}
\end{figure}
In Fig. \ref{m01} the time evolution of the B-luminosity for an Sc
galaxy with different interactions affecting its SFR after 6 Gyr of
undisturbed evolution is shown. The flat curve shows the evolution of
an undisturbed Sc galaxy. As a result of its star formation rate
decreasing slowly as the gas content decreases, the B-band luminosity
of the undisturbed Sc galaxy model (dot-dashed) also decreases very
little after its maximum is reached at $\sim 2$ Gyr of evolution.
After 12 Gyr of undisturbed evolution the average ${\rm \langle
  M_B\rangle \sim -18.3~mag}$ of Virgo Sc galaxies is reached.  If the
star formation is truncated at 6 Gyr as shown by the solid curve, the
B-luminosity decreases appreciably and only amounts to ${\rm M_B}=
-16.9$ mag at an age of 12 Gyr.  Bursts of various strengths, consuming
30 and 70\% of the remaining gas reservoir, lead to a strong and rapid
luminosity increase to ${\rm M_B}= -19.5$ and ${\rm M_B}= -20.3$ mag for
the weak and strong burst cases (dotted and dashed lines),
respectively. About $5\cdot 10^8$yr after the onset of the burst,
systems start fading again and 2 Gyr (4 Gyr) after the weak (strong)
burst, the post-burst galaxy becomes fainter than the undisturbed one.
If some residual star formation is assumed at a constant rate of ${\rm
  1.5 ~M_{\odot}yr^{-1}}$ after the burst (long dashed line), the
fading gets weaker.

\begin{figure}[h]
\includegraphics[width=\columnwidth,angle=0]{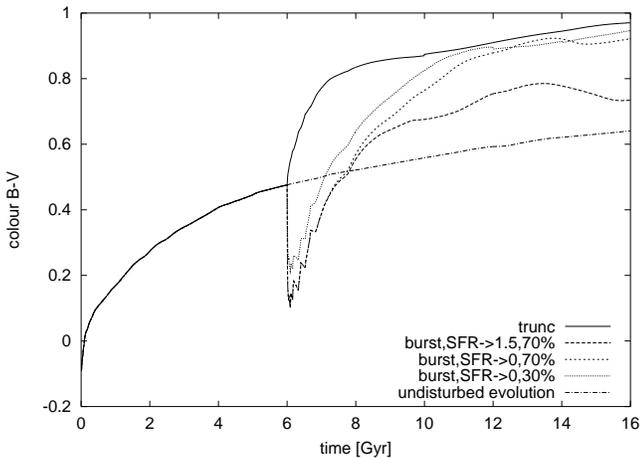}
\caption{Time evolution of the ${\rm B-V}$ color for the same models as in Fig.\ref{m01} }
\label{m02}
\end{figure}
Fig. \ref{m02} shows the color evolution in ${\rm B-V}$ for the same
models as in Fig. \ref{m01}. The flat dot-dashed curve is the
undisturbed Sc galaxy again. It describes the successive reddening of
its aging stellar population. With the star formation rate we assumed,
the Sc model, after 12 Gyr of undisturbed evolution, indeed reaches
the average ${\rm B-V}= 0.59$ of Sc galaxies in the RC3 (Buta \etal
1994). If star formation is truncated after 6 Gyr of evolution (solid
line) a strong and very sudden reddening results due to the massive
stars dying out very fast. 2 Gyr after star formation truncation,
${\rm B-V}=0.84$ is reached, whereafter the further reddening is much
slower and reaches ${\rm B-V}=0.91$ at 12 Gyr.  If a burst occurs
after 6 Gyr of evolution, colors rapidly get very blue, the difference
between a strong burst (${\rm B-V}=0.1$, dashed line) and a weak one
(${\rm B-V}=0.2$, dotted line) not being large. After the burst, ${\rm
  B-V}$ reddens again, though more slowly than in the truncation case.
The ${\rm B-V}$ of the undisturbed model is reached 1.5 Gyr after a
weak and 1.8 Gyr after the strong burst. Thereafter the reddening
continues until, at an age around 12 Gyr, ${\rm B-V}$ is almost as red
as in the truncation model. If star formation does not go to zero
after the burst but continues at some low constant level, ${\rm B-V}$
stays significantly bluer ${\rm (B-V) \sim 0.75}$.  Note that while
burst models with ${\rm \psi_f=0}$ and star formation truncation
models cannot be distinguished in ${\rm B-V}$ any more at times $\ge
5$ Gyr after the event, they are still very well distinguished in
terms of luminosity.

Here, as an example, we only present the B-luminosity and ${\rm
  B-V}$-color evolution. Our models, however, comprise all optical and
NIR luminosities and colors from U through K and they all behave in
their specific ways in burst and truncation models.

For undisturbed galaxies of all types Sa, Sb, Sc and Sd, of course,
star formation rates were chosen as to provide agreement of all colors
from ${\rm U-B}$ through ${\rm V-K}$ at an evolutionary age of $\sim
12$ Gyr with their respective average values as observed for nearby
galaxy samples.

Because of their initially higher and more rapidly decreasing star
formation rates, the B-band luminosity evolution of Sa and Sb models
has a more pronounced maximum at early times whereafter it shows a
stronger fading than the Sc model. As well, the ${\rm B-V}$ colors of
Sa and Sb models become redder, almost from the very beginning, than
those of the Sc model shown here. This is due to the larger fraction
of stars formed at early times.

The constant star formation Sd model has almost constant B-luminosity
from $\le 2$ to $\ge 12$ Gyr but still shows some reddening, albeit
weaker than the Sc model, due to the accumulation of low mass stars.

Note that the gas content drops at very different rates in different
types of models, very fast in Sa with $(\textrm{G}/\textrm{M})_{\rm
  12~Gyr}=0.07$ and pretty slowly in Sd models with
$(\textrm{G}/\textrm{M})_{\rm 12~Gyr}=0.65$. These global gas contents
are also in broad agreement with typical gas-to-total mass ratios of
field galaxies if the HI beyond their optical radius is taken into
account. Without significant accretion from outside these evolving gas
masses set upper limits to the star formation in bursts. In our
definition of the burst strength in terms of consumed gas fraction, a
70\% (strong) burst in an old (9 Gyr) and hence already gas-poor Sa
galaxy only increases its stellar mass by 0.8\% while a 70\% burst in
a much more gas-rich 3 Gyr young Sc model would increase its stellar
mass by almost a factor of 2.

Table \ref{btab} compares the different definitions of the burst strength parameters
for bursts after 6 Gyr of undisturbed evolution.

\begin{table}
\begin{center} 
\begin{tabular}{|c|c|c|} 
\hline 
Hubble Type&b& ${\rm b_*}$\\
\hline 
Sa  & 30\% & 6\% \\
    & 70\% & 16\% \\
Sb  & 30\% & 13\% \\
    & 70\% & 27\% \\
Sc  & 30\% & 29\% \\
    & 70\% & 52\% \\
\hline 
\end{tabular} 
\caption{Comparison of the two different burststrength definitions ${\rm b:=\frac{\Delta
G}{G(t_{burst})}}$ and ${\rm b_*:=\frac{\Delta S}{S(t_0)}}$ for galaxies bursting at 6 Gyr.} 
\label{btab} 
\end{center} 
\end{table}

\section{Comparison with S0 Galaxy Data}

As a first application of our models we study the origin of the S0 galaxy population in
cluster environments. The conventional model of S0 galaxy evolution is a high SFR at the
beginning followed by a strong decrease of star formation activity (cf. Sandage 1986,
Bruzual \& Charlot 1993, Guiderdoni \& Rocca -- Volmerange 1987). In this picture, S0
galaxies are old and have evolved passively most of the time. Recent observations, however,
show that the fraction of S0 galaxies in clusters evolves strongly with redshift. We assume
that interactions among infalling galaxies and with the cluster environment trigger the
evolution. We take the turn-over ${\rm M_B}$ of the Gauss-fit determined by Sandage
\etal (1985) to the S0 galaxy luminosity function in Virgo (adjusted to our choice of ${\rm
H_0=65}$) as  our reference for the local average S0 luminosity ${\rm \langle M_B\rangle}$.
Note that the completeness limit for this sample occurs about 5 mag fainter than ${\rm
\langle M_B\rangle}$. The average colors of local S0 galaxies over the same luminosity
range are taken from Buta \etal (1994) and Bower \etal (1992). Table \ref{tab1} shows the
data we use for the local S0 galaxy population, in our comparison with models of
interacting and disturbed spiral galaxies described above.

\begin{table}[h]
\begin{center} 
\begin{tabular}{|c|c|c|} 
\hline 
${\rm \langle M_B\rangle}$&$-18.55 \pm 1.5~$mag& Sandage et al. 1985 \\ 
${\rm B-V}$  &$0.90 \pm 0.060~$mag& Buta et al. 1994\\
${\rm U-B}$  &$0.46 \pm 0.097~$mag& Buta et al. 1994\\ 
${\rm V-K}$  &$3.15 \pm 0.100~$mag& Bower et al. 1992 \\ 
\hline 
\end{tabular} 
\caption{Average S0 galaxy data with $1\sigma$ dispersions.} 
\label{tab1} 
\end{center} 
\end{table}

The aim is to find out which effects on the star formation rates in
which types of galaxies and at what times do produce colors and
luminosities by today in agreement with S0 galaxy data. This kind of
investigation is meant to complement dynamical models for the
morphological transformation of galaxies in cluster environments
(Moore \etal 1996, 1998).

It was seen from the models alone (Sect.3) that while in terms of
${\rm B-V}$ colors at $\sim 12$ Gyr several very different models
looked identical, they considerably split up in terms of B-band
luminosity. We therefore chose to present here our comparison of
models and S0 data in terms of color-magnitude diagrams.

\subsection{Starburst Models}
\begin{figure}
\includegraphics[width=\columnwidth,angle=0]{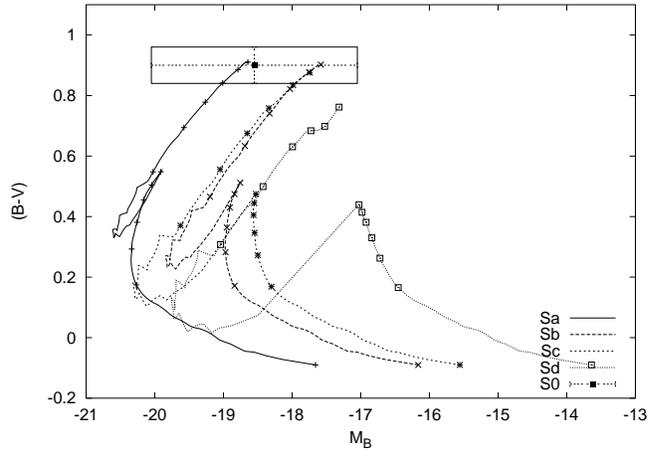}
\caption{Color-magnitude diagram for the evolution of spiral galaxies
  with strong bursts starting at an age of 6 Gyr. The symbols mark
  timesteps of 1 Gyr.}
\label{s01}
\end{figure}

In Fig. \ref{s01} we show the evolution of model galaxies of types Sa,
Sb, Sc, and Sd with strong starbursts starting at an age of 6 Gyr in a
color-magnitude diagram ${\rm (B-V)}$ vs.  ${\rm M_B}$. The
evolutionary tracks start in the lower part of the diagram. Timesteps
of 1 Gyr are shown by the symbols on each track.  As shown in Fig.
\ref{m01} and \ref{m02} for undisturbed Sc galaxies, all models get
brighter and redder, and this evolution is fastest in the earliest
stages and gets slower later on. At an age of 6 Gyr all models are
assumed to have strong starbursts. Hence, the model galaxies get bluer
and brighter very fast and then, with falling SFR, they get redder and
fainter again. The strengths of the changes in color and luminosity
during the burst depend on galaxy type.  The Sd model, with a lot of
available gas, passes through a big loop in the color-magnitude
diagram, while the Sa model, with much less gas, only shows a small
loop. Remember, all models have a burst of the same strength in our
definition, i.e. they transform 70\% of their available gas mass into
new stars.

The box in the upper left of the color-magnitude diagram indicates the $1\sigma$ range
around the average luminosity and color of the local S0 galaxy population (see also
Tab.~\ref{tab1}). The endpoints of the Sa, Sb, and Sc evolutionary tracks, at an age of 12
Gyr, lie within the observed 1$\sigma$ range of local S0 galaxies. Only the Sd model is too
blue to reach to within $1\sigma$ of the photometric properties of S0s.

The following figures only show the end points of galaxy evolution at
12 Gyr. In all plots the sequence of undisturbed galaxies is shown for
reference. Lines connect the various galaxy types with
the same interaction history (from left to right: Sa, Sb, Sc, and Sd).
The box in the upper left again shows the $1\sigma$ range of the local S0
population.

\begin{figure}
\includegraphics[width=\columnwidth,angle=0]{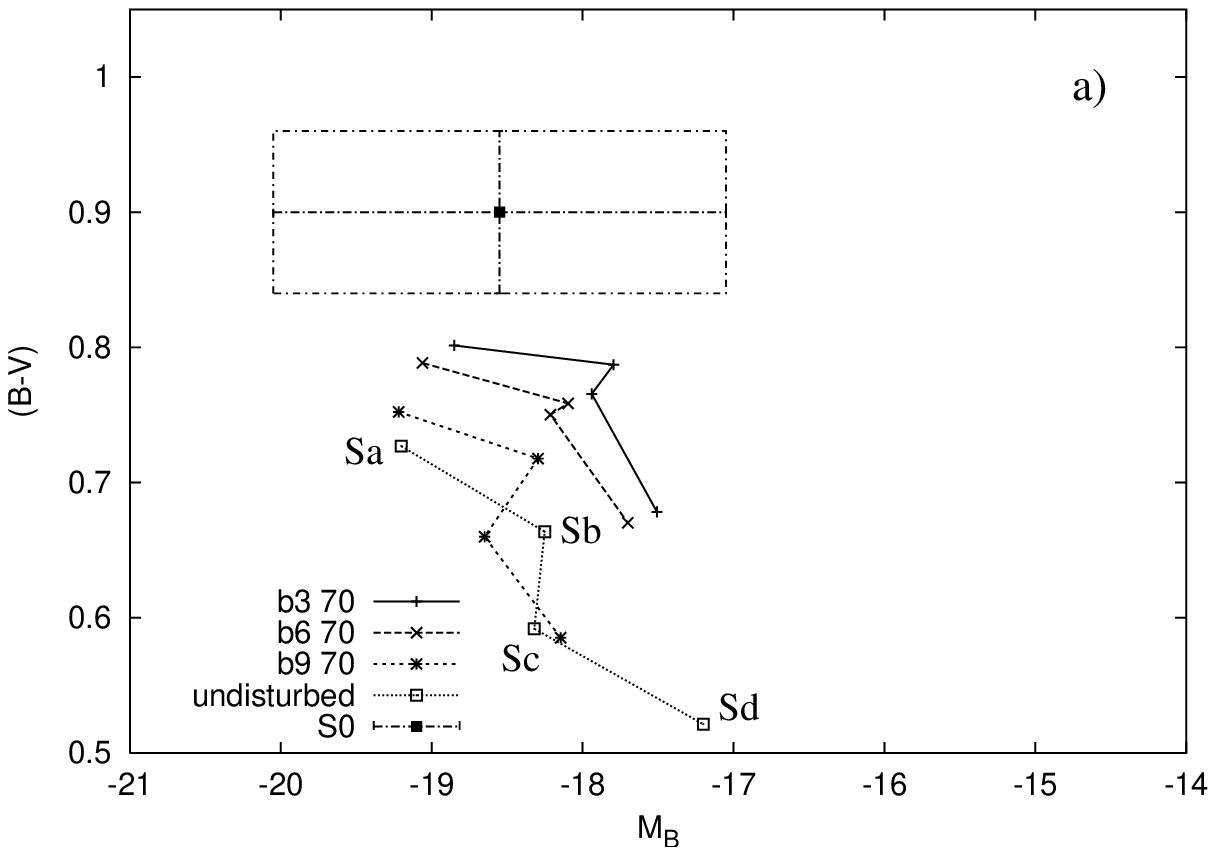}
\includegraphics[width=\columnwidth,angle=0]{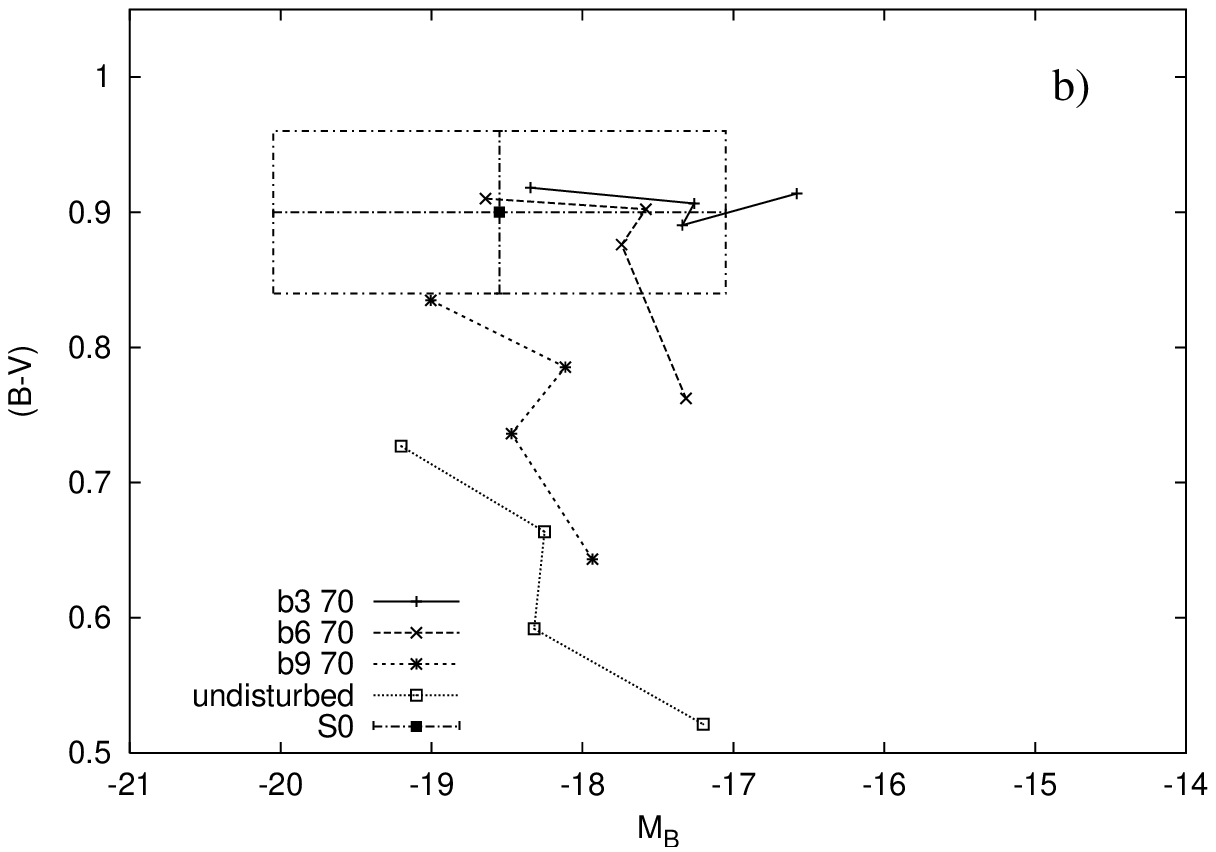}
\caption{Difference between burst models with and without residual SFR
  ${\rm \psi_f=1.5~M_{\odot}yr^{-1}}$ (a) and ${\rm \psi_f=0}$ (b).}
\label{s03}
\end{figure}

Figs \ref{s03}a and b show for starburst models the effect of a
residual ongoing SFR (${\rm \psi_f=1.5 ~M_{\odot}yr^{-1}}$) after the
burst (Fig. 4a) as compared to starburst models with no ongoing star
formation (4b). In both figures strong (70\%) burst at three different
times of evolution (3, 6, 9 Gyr) are plotted. The differences are very
clear. No model with ongoing low level star formation reaches the
luminosity {\bf and} color range ($1\sigma$) of local S0 galaxies. All
model galaxies are too blue. This is the consequence of the ongoing
star formation, as the comparison with Fig \ref{s03}b shows. Without
ongoing star formation after the burst several model galaxies reach
the $1\sigma$ range of local S0 galaxies in luminosity and color. Only
the galaxies which have their burst at an evolutionary age of 9 Gyr
(this means only 3 Gyr ago) are (still) too blue. The Sa galaxy,
however, even after a burst only 3 Gyr ago is already very close to
the colors of S0 galaxies.

We conclude that {\bf all spiral galaxies with bursts at ages of 3 or
  6 Gyr, except for the Sds, reach both the luminosity and color
  ranges of S0s by today}, and this is not only true for ${\rm M_B}$
and (B$-$V) displayed here but for all other colors available. {\bf
  The Sd galaxies are either too faint (burst at 3 Gyr) or too blue
  (bursts at 6 or 9 Gyr) to resemble by today the average photometric
  properties of S0 galaxies. An essential prerequisite for
  post-starburst models to reach the red colors of S0 galaxies is a
  complete extinction of SF.}

\subsection{Star Formation Truncation Models}

\begin{figure}
\includegraphics[width=\columnwidth,angle=0]{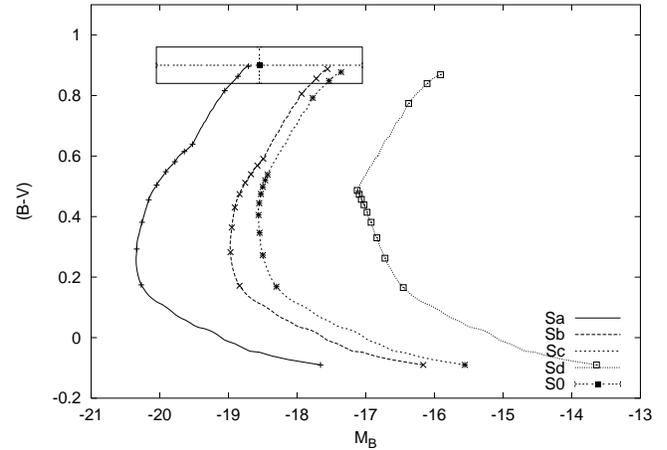}
\caption{Color-magnitude diagram for the evolution of spiral galaxies
  with star formation truncation after 9 Gyr of undisturbed evolution.
  The symbols mark timesteps of 1 Gyr. }
\label{s02}
\end{figure}

Another interaction scenario is shown in Fig.~\ref{s02}. Here the SFR
is truncated after 9 Gyr of undisturbed evolution. It is a
color-magnitude diagram like Fig.~\ref{s01} with the same scale. So
the models follow the tracks from the bottom to the top, ending at 12
Gyr with symbols again marking timesteps of 1 Gyr. Like in
Fig.~\ref{s01}, the Sa, Sb and Sc models -- now with SF truncation --
reach the observed range of local S0 galaxies in the upper left, while
the Sd model again misses the S0 region. In this scenario, however,
the Sd model fails because its luminosity is too low while its color
now is red enough, in contrast to the burst scenario. It is also seen
that the effect of the SF truncation (the break in the curve) is
smaller for the Sa model and gets stronger towards the Sd model.  The
explanation is that the SFR in an Sa galaxy is already low at this
time of evolution and the galaxy is dominated by old stars and rather
red already, so a truncation of the SFR does not affect the galaxy too
much. The Sd galaxy, on the other hand, is actively forming stars even
at an age of 9 Gyr. A truncation of its SFR causes a sudden decrease
in the number of young hot stars, and the effect on the galaxy in
terms of B-band luminosity and color is stronger.

\begin{figure}
\includegraphics[width=\columnwidth,angle=0]{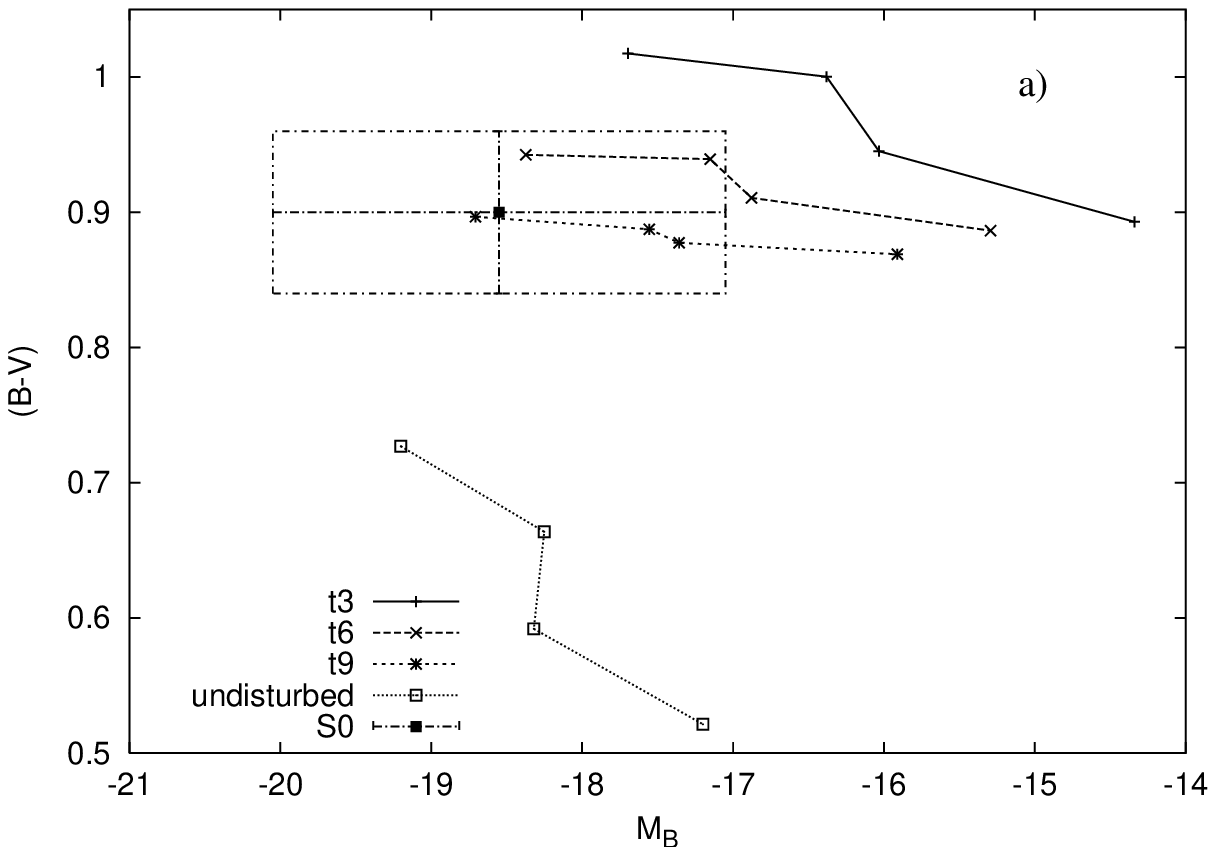}
\includegraphics[width=\columnwidth,angle=0]{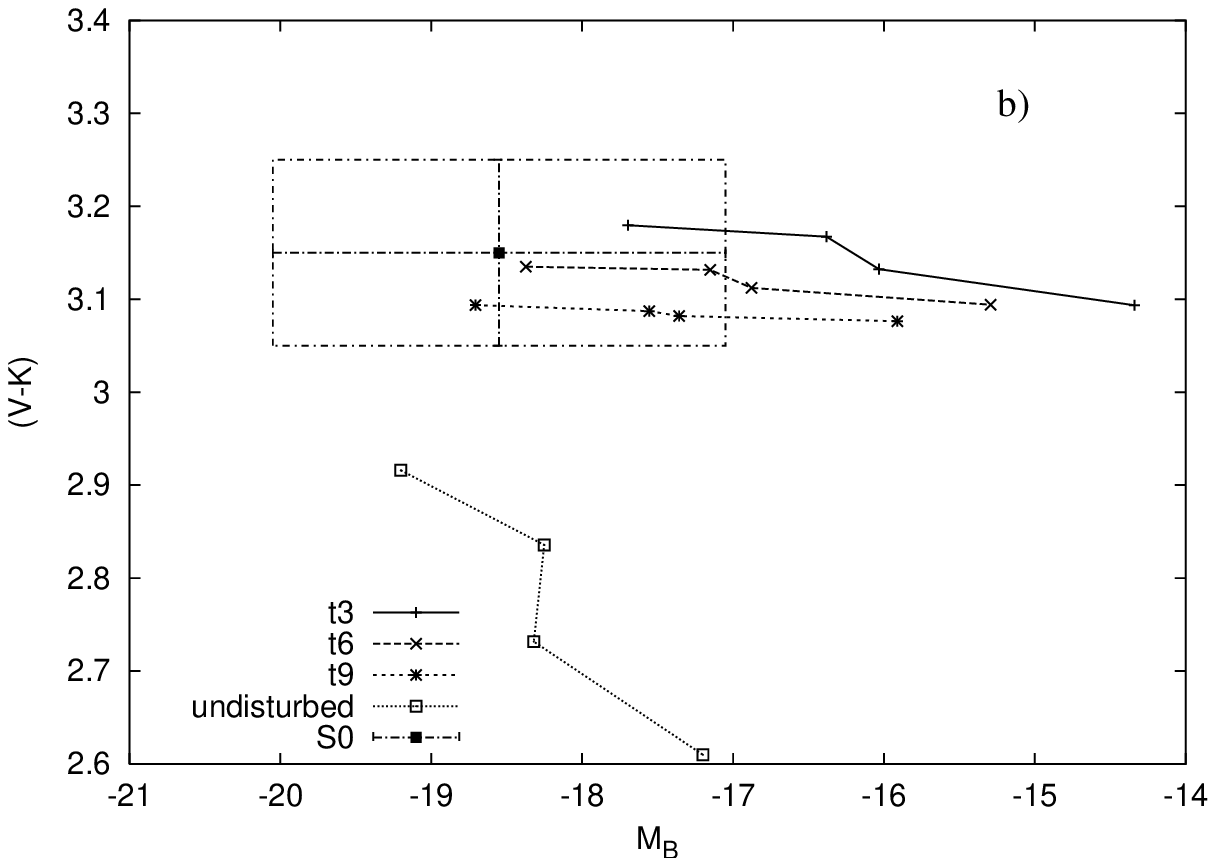}
\caption{Effect on galaxy colors and luminosities at $\sim 12$ Gyr of SF truncation in various spiral types
(connected by lines to guide the eye) occurring at various evolutionary ages. The two Figs show different
colors ${\rm (B-V)}$ (a), and ${\rm (V-K)}$ (b).}
\label{s04}
\end{figure}

The next two Figures, Fig. 6a) and b) show SF truncation scenarios, occurring at 3, 6, and
9 Gyr of evolution in different spiral galaxy types. In Fig. 6a) ${\rm B-V}$, and in
6b) ${\rm V-K}$ colors are plotted against ${\rm M_B}$. Other colors like ${\rm U-B}$ or
${\rm U-V}$ do not give any additional information and therefore are not showen. If the SFR is truncated
at early times (3 Gyr) the colors become too red (Sa, Sb) and the luminosities to faint
(Sb, Sc, Sd) by today. Is the SFR truncated at 6 or 9 Gyr, most models reach the $1\sigma$
range of local S0 galaxies.  Only Sd galaxies are an exception, they are too faint because
they did not build enough stars to get as bright as an average S0 galaxy.  In Fig. 6b) the
conclusions for models which have a SF truncation after 6 or 9 Gyr are equal to those
described above. But in comparison to a) the Sa galaxy is not too red in ${\rm V-K}$ in
case of SF truncation at 3 Gyr. This is due to the fact that the old stars are dominating
${\rm V-K}$ while in ${\rm U-B}$ or ${\rm B-V}$ young hot stars dominate.  So the
truncation of the SFR is less visible in ${\rm V-K}$ as compared to ${\rm U-B}$ or ${\rm
B-V}$.

Hence, the ${\rm (B-V)~vs~M_B}$ diagram already contained all the
information. The inclusion of U- and K-bands does not lead to any
further restriction of the manifold of S0 galaxy progenitor models. We
conclude that {\bf Sa models with SF truncation after 6 or 9 Gyr of
  evolution fall well into the color and luminosity range of
  present-day S0 galaxies. Sb and Sc galaxies with SF truncation at 6
  or 9 Gyr and at 9 Gyr, respectively, may be the progenitor of lower
  than average luminosity S0s. Late-type Sd spirals with SF truncation
  at any time are too faint.}

\subsection{Effects of the Burst Strength}

\begin{figure}
\includegraphics[width=\columnwidth,angle=0]{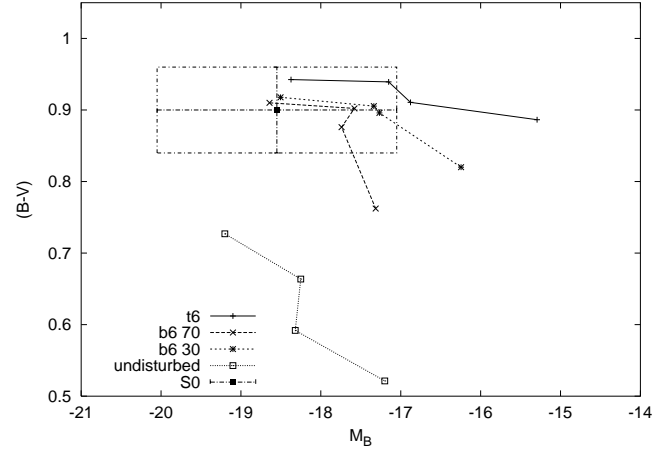}
\caption{Comparison of models with various burst strengths, $b=0.7$ and $b=0.3$ (${\rm \psi_f=0}$) and with SF truncation, all occurring at an age of 6 Gyr.}
\label{s08}
\end{figure}

In Fig.~\ref{s08} we compare different burst strengths and SF
truncation (which can also be understood as a b=0 model) on galaxies
which have their SFR disturbed at 6 Gyr. The burst strengths are 70\%
and 30\% without ongoing star formation thereafter. Again, the Sd
galaxies lie all outside of the S0 range. Their ${\rm B-V}$ colors and
B-luminosities are strongly dependent on burst strength. All other
spiral galaxy types lie much closer to each other, in particular the
differences between the strong and weak bursts are small. All these
galaxies reach the $1\sigma$ range of today's S0s, only the truncated
Sc galaxy marginally misses the S0 luminosity range. Hence, our models
show that {\bf the burst strength is not an important factor in the
  transformation of a spiral progenitor into an S0 galaxy. The
  important factor is the SFR which has go to zero by today, either by
  a direct truncation, or after a burst.}

\subsection{Merger-induced Starbursts}

\begin{figure}
\includegraphics[width=\columnwidth,angle=0]{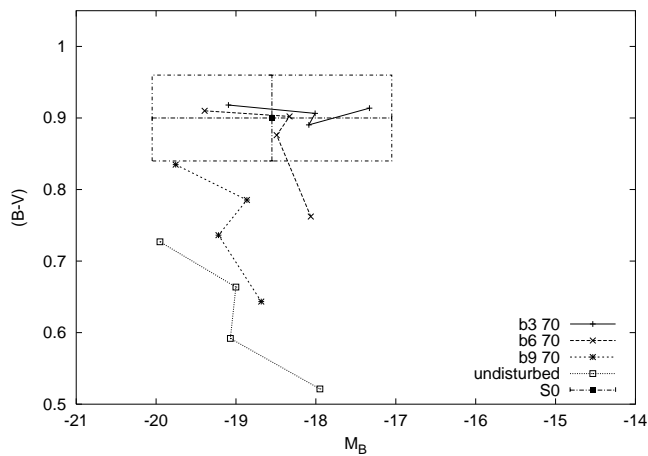}
\caption{Color-magnitude diagram for Spiral-Spiral merger models with strong starbursts and
SFR=0 after the burst}
\label{s07}
\end{figure}

Last, we study the merger scenario. As mentioned before, the high velocity dispersions of
galaxies in local rich clusters clearly are unfavorable for galaxy mergers. In infalling
galaxy groups, still today, and during early stages of cluster formation, however, galaxy
merging may play or have played a role. We restrict ourselves to mergers of equal spiral
types, and explore the starburst that may be triggered in the course of merging. These
types of mergers double the mass -- in stars as well as in gas -- of our model galaxies,
and also double the luminosity of the merger remnants, corresponding to a brightening by
0.75 mag.  The colors do not change with respect to the simple starburst models discussed
above. Fig.~\ref{s07} shows galaxies which merged at 3, 6, and 9 Gyr and experienced a
strong starburst at this occasion. Comparison with Fig.~\ref{s03}b for starbursts in
single galaxies shows this difference of 0.75 mag in luminosity. So, like all others, the
Sd-Sd merger remnants are brighter and in the case of a burst at 3 Gyr they now can reach
the faint part of the S0 luminosity distribution. The Sa galaxies are pushed by mergers
into the luminosity range of bright S0s, while in the non-merger starburst scenario they
only reached the average S0 luminosity. {\bf Mergers of different normal galaxy types
and mergers between a normal and a dwarf galaxy never get brighter than an Sa-Sa merger.
Hence, the only ways to produce brighter  than average S0s are either early-type spiral
merger with starbursts induced at 3 or 6 Gyr, or multiple mergers. Sb -- Sb and Sc -- Sc
mergers at 3 or 6 Gyr can well account for the average luminosity S0 population and even
early mergers (at 3 Gyr) of late-type Sd galaxies do reach the observed color and
luminosity range of present-day S0s. SFRs have to go to zero after the bursts in all cases
to obtain red enough colors}. Sd -- Sd mergers at 6 Gyr and all types of late mergers (at
${\rm t \geq 9}$ Gyr) remain too blue.

\subsection{Discussion} 

Note that we show the $1\sigma$ range of the local S0 galaxy population in all of our plots
and so our models also describes average luminosity progenitor galaxies. Hence, some of the
models marginally missing the local S0 $1\sigma$ box in therms of luminosity may still
describe valid ways to transform spirals into S0s.
Within infalling groups, it is, of course, also possible that more than two galaxies may
merge during their infall into a cluster, giving rise to a bright S0 galaxy.

The results of our photometric study are in good agreement e.g. with Jones \etal
(2000)  who also quote star formation truncation as the dominant transformation process
from field  spirals to cluster S0s from analysis of spectral features and HST morphologies 
of S0 galaxies in clusters at redshift 0.3 -- 0.6.  Our conclusions confirm their results
obtained for the brighter part of the S0 population accessible to spectral analyses and
goes beyond that study in the inclusion of the fainter part of the local S0 population. On
the other hand, more than 40\% of the S0 galaxies in the Coma cluster are found to show
evidence for star formation in their central regions during the last $\sim 5$ Gyr by
Poggianti et al. (2001b) from spectroscopic analyses, in particular at fainter
luminosities, pointing towards a starburst  prior the star formation truncation,
complementing our findings that most pure star formation truncation models have remnants
fainter than average S0 galaxies. 

Poggianti \etal (2001a) find that the fraction of present-day luminos galaxies that had
significant SF between z=2 and z=0.35 is higher than for lower luminosity galaxies. This as
well, is in agreement with ourresults that luminous S0s must have had starbursts earlier in
their evolution while lower luminous ones may also result from pure SF truncation.

We conclude that two scenarious may be responsible for the transformation of field spiral
galaxies colors and luminosities into those of S0s. Starbursts and possibly even
spiral-spiral mergers seem necessary for the luminous part of the cluster S0 population
while SF truncation without any significant preceding enhancement may explain the lower
luminous S0s.    

Our finding that S0 galaxies in clusters may result from starbursts in spirals
at all ages below about 9 Gyr is not surprising in the sense that more recent
bursts may still show features of postburst galaxies and hence not be
classified as S0s. Our result that SF truncation, on the other hand, if
occuring too early, would make galaxies redder than observed for today's S0s,
may have to do with the fact that the hot and dense ICM in the centers of
galaxy clusters thought to be responsible for sweeping the gas out of the
infalling spirals needs time to pile up and may not be efficient in truncating
SF before ${\rm z \sim 0.5~.~.~.~1}$.

\section{Conclusions and Outlook}
The fraction of S0 galaxies in clusters is observed to significantly
depend on the redshift of the cluster. In nearby clusters the fraction
of S0 type galaxies is about 40 \%, whereas in clusters up to z=0.5 it
decreases by a factor 3 to 4. Various interaction processes have been
discussed that might act to transform spirals from the field galaxy
population falling into a cluster into passive S0 galaxies. Starbursts
may be triggered by shock waves caused in the ISM when a galaxy falls
into the hot dense cluster ICM or by mergers within groups of
infalling galaxies. Star formation truncation, on the other hand, may
occur in galaxies which lose their gas by tidal interaction or by ram
pressure stripping when falling into a cluster.

In this paper we study the origin of S0 galaxies in clusters by means
of photometric evolutionary synthesis. Our models calculate the time
evolution of global luminosities U, ..., K and colors, as well as of
the stellar and gaseous content of various types of undisturbed spiral
galaxies Sa -- Sd. These are parameterized by their respective
appropriate star formation histories and models are normalized to
reproduce, after $\sim 12$ Gyr of evolution, the average B-band
luminosity locally observed for the respective spiral type. We
computed a grid of models for spiral galaxies falling into a cluster
and having their SFRs affected by starbursts and/or star formation
truncation occurring at various evolutionary times. The luminosities
and colors of the disturbed galaxies at an age of $\sim 12$ Gyr are
compared to the average observed photometric properties (B-band
luminosity, (B--V), (U--B), and (V--K) colors) of local S0 galaxies.
Our aim was to find out which of the disturbed models reach the color
and luminosity range of S0 galaxies in the Virgo and Coma clusters
and, hence, to constrain the manifold of transformation processes.

We find that many of our disturbed galaxy models do evolve into the
color and luminosity range of local S0 galaxies.

\begin{itemize}
\item[-] The strongest constraint is that the SFR after the
  interaction has to go to zero. Models with even low-level ongoing SF
  after a burst or after SF truncation remain too blue. If before the
  total SF truncation a burst occurs, even a strong one, or not, is of
  minor importance.
\item[-] All spiral galaxy types Sa -- Sc with starbursts that occurred
  more than 3 Gyr ago reach S0 galaxy colors and luminosities by
  today. Sd galaxies with early or late bursts are too faint or too
  blue, respectively.
\item[-] Sa models with SF truncation at ages between 6 and 9 Gyr well
  account for the photometric properties of present-day S0s. Sb
  galaxies with SF truncation at 6 or 9 Gyr and Sc galaxies with SF
  truncation at 9 Gyr may account for the lower luminosity part of the
  S0 population. Sd galaxies with SF truncation at any time are too
  faint.
\item[-] The strength of a starburst is not an important factor for
  the development of photometric resemblance to S0 galaxies as long as
  the SFR after the burst declines enough. 
\item[-] We find that photometric constraints leave two scenarios viable 
  for the transformation of field spirals into cluster S0s: while low 
  luminosity S0s may be the result of simple SF truncation in spirals 
  earlier than Sd, the higher luminosity S0s require starbursts or even 
  mergers prior to SF extinction.   
\item[-] Our results imply a difference in the cosmological epoch of 
  the transformation of field spirals to cluster S0s in the two scenarios. 
  Starbursts occuring in spirals at all ages lower than about 9 Gyr leave 
  remnants with average S0 colors while SF truncation should not occur at 
  epochs earlier than ${\rm z \sim 0.5,~.~.~.~1}$ in order not to make 
  the remnants redder than present-day S0s. This may be related to the 
  fact that the hot and dense ICM thought to be responsible for sweeping 
  the gas out of the infalling spirals and truncating SF may not yet be 
  efficient at earlier epochs. 
\end{itemize}

Our photometric evolutionary synthesis approach is meant to complement
dynamical simulations and studies of spectral features. A more
detailed analysis of the redshift evolution of the properties and
relative numbers of S0 galaxies in clusters will be the next step in
our attempt to constrain possible evolutionary paths from the
spiral-rich field galaxy population to the S0-rich local cluster
population. Spectral modeling including evolutionary and cosmological
corrections will have to be compared to observations throughout
clusters of various richness, degree of relaxation, and redshift.

\begin{acknowledgements}
We thank our referee, Dr. Bianca Poggianti for a very prompt, insightful
and constructive report.
\end{acknowledgements}

\end{document}